\documentclass[reprint,superscriptaddress,showpacs,amsmath,amssymb,prb]{revtex4-1}

\usepackage{graphicx}
\usepackage{wrapfig}
\usepackage{epsfig}
\graphicspath{ {./images/} }
\usepackage{dcolumn}
\usepackage{bm}
\usepackage{amsmath}
\usepackage{gensymb}
\usepackage{epstopdf}
\usepackage{cuted}
\usepackage{mathtools}
\usepackage{upgreek}
\usepackage{natbib}
\usepackage{hyperref}
\usepackage[mathlines]{lineno}

\begin{document}

\preprint{APS/123-QED}

\title{Crystal Symmetry Lowering in Chiral Multiferroic Ba$_3$TaFe$_3$Si$_2$O$_{14}$ observed by X-Ray Magnetic Scattering}

\author{M. Ramakrishnan}
 \email{mahesh.ramakrishnan@psi.ch}
 \affiliation{Swiss Light Source, Paul Scherrer Institut, 5232 Villigen PSI, Switzerland}
\author{Y. Joly}
 \affiliation{Universit\'{e} Grenoble Alpes, Institut NEEL, F-38042 Grenoble, France}
 \affiliation{CNRS, Institut NEEL, F-38042 Grenoble, France}
\author{Y. W. Windsor}
 \altaffiliation{Present address: Department of Physical Chemistry, Fritz-Haber-Institut of the Max Planck Society, Faradayweg 4-6, Berlin 14915, Germany}
 \affiliation{Swiss Light Source, Paul Scherrer Institut, 5232 Villigen PSI, Switzerland}
\author{L. Rettig}
 \altaffiliation{Present address: Department of Physical Chemistry, Fritz-Haber-Institut of the Max Planck Society, Faradayweg 4-6, Berlin 14915, Germany}
 \affiliation{Swiss Light Source, Paul Scherrer Institut, 5232 Villigen PSI, Switzerland}
 \author{A. Alberca}
 \affiliation{Swiss Light Source, Paul Scherrer Institut, 5232 Villigen PSI, Switzerland}
 \affiliation{University of Fribourg, Department of Physics and Fribourg Centre for Nanomaterials, Chemin du Musee 3, CH-1700 Fribourg, Switzerland}
\author{E. M. Bothschafter}
 \affiliation{Swiss Light Source, Paul Scherrer Institut, 5232 Villigen PSI, Switzerland}
\author{P. Lejay}
 \affiliation{Universit\'{e} Grenoble Alpes, Institut NEEL, F-38042 Grenoble, France}
 \affiliation{CNRS, Institut NEEL, F-38042 Grenoble, France}
\author{R. Ballou}
 \affiliation{Universit\'{e} Grenoble Alpes, Institut NEEL, F-38042 Grenoble, France}
 \affiliation{CNRS, Institut NEEL, F-38042 Grenoble, France}
\author{V. Simonet}
 \affiliation{Universit\'{e} Grenoble Alpes, Institut NEEL, F-38042 Grenoble, France}
 \affiliation{CNRS, Institut NEEL, F-38042 Grenoble, France}
\author{V. Scagnoli}
 \email{valerio.scagnoli@psi.ch}
 \affiliation{Laboratory for Mesoscopic Systems, Department of Materials, ETH Zurich, 8093 Zurich, Switzerland.}
 \affiliation{Laboratory for Multiscale Materials Experiments, Paul Scherrer Institut, 5232 Villigen PSI, Switzerland}
\author{U. Staub}
 \email{urs.staub@psi.ch}
 \affiliation{Swiss Light Source, Paul Scherrer Institut, 5232 Villigen PSI, Switzerland}
 
\begin{abstract}
Chiral multiferroic langasites have attracted attention due to their doubly-chiral magnetic ground state within an enantiomorphic crystal. We report on a detailed resonant soft X-ray diffraction study of the multiferroic Ba$_3$TaFe$_3$Si$_2$O$_{14}$ at the Fe $L_{2,3}$ and oxygen $K$ edges. Below $T_N$ ($\approx27K$) we observe the satellite reflections $(0,0,\tau)$, $(0,0,2\tau)$, $(0,0,3\tau)$ and $(0,0,1-3\tau)$ where $\tau \approx 0.140 \pm 0.001$. The dependence of the scattering intensity on X-ray polarization and azimuthal angle indicate that the odd harmonics are dominated by the out-of-plane ($\mathbf{\hat{c}}$-axis) magnetic dipole while the $(0,0,2\tau)$ originates from the electron density distortions accompanying magnetic order. We observe dissimilar energy dependences of the diffraction intensity of the purely magnetic odd-harmonic satellites at the Fe $L_3$ edge. Utilizing first-principles calculations, we show that this is a consequence of the loss of threefold crystal symmetry in the multiferroic phase.
\end{abstract}


\maketitle


\section{Introduction}

Electric field control of magnetism has been one of the major scientific and technological goal that has risen to the forefront of condensed matter research in the last few decades \cite{Fiebig2005,Spaldin2005,Eerenstein2006,Cheong2007}. Very few systems simultaneously possess multiple ferroic orders at room temperature, and the materials where these orders are strongly coupled form an even smaller subset \cite{Hill2000}. At the heart of such multiferroic behavior, there exists a complex interplay of charge, spin, lattice and orbital degrees of freedom \cite{Khomskii2006}. Many systems featuring strong magnetoelectric interactions are non-collinear antiferromagnets\cite{Tokura2014}, where the existence of strong spin-lattice coupling in the ground state effectively lifts inversion symmetry, resulting in a net electric polarization \cite{Kenzelmann2005,Kimura2003}. Understanding of the lattice structure and magnetic symmetry are hence, of utmost importance to determine the nature of magnetoelectric interactions in these materials. \par
 
\begin{figure}[h] 
\centering
\includegraphics[width=0.5\textwidth]{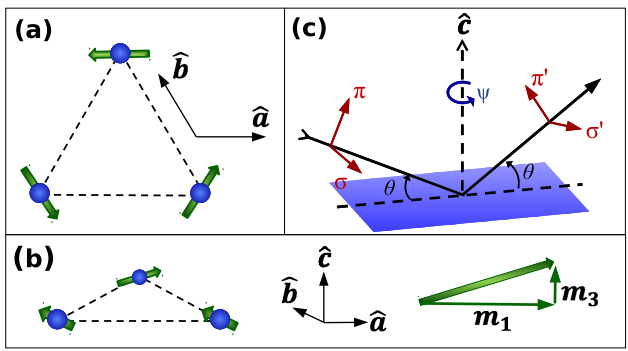}
\caption{Schematic representation of the spin structure within the triangular units of BTFS, when viewed (a) along $\mathbf{\hat{c}}$-axis and (b) at an angle to the \emph{ab} plane, with a schematic representation of the components of the magnetic moment along the Cartesian axes. The threefold crystal axis is along $\mathbf{\hat{c}}$, whereas the local twofold axes lie along $\mathbf{\hat{a}}$ and $\mathbf{\hat{b}}$ directions. (c) Experimental geometry and definition of the photon polarization states. The azimuthal rotation angle $\Psi$ is defined as zero when the crystallographic $\mathbf{\hat{a}}$- and $\mathbf{\hat{c}}$-axis are in the scattering plane}
\label{setup}
\end{figure}

Several compounds belonging to the langasite family have been studied in recent years for their piezoelectric and nonlinear optical properties \cite{Marty2008,Marty2010,Lyubutin2011,Loire2011,Pikin2012}. They have a general formula $A_{3}BC_{3}D_{2}O_{14}$, and crystallize in the noncentrosymmetric chiral space group $P321$ at room temperature. Crystals of Ba$_3$TaFe$_3$Si$_2$O$_{14}$ (BTFS) (and its isostructural compound Ba$_3$NbFe$_3$Si$_2$O$_{14}$ (BNFS)) grow in an enantiopure phase with the magnetic $Fe^{3+}$ ions sitting at tetrahedral sites in the structure (space group $P321$), forming a network of triangular units in the basal (\emph{ab}) plane. Below $T_N = 27 K$, the spins of the $Fe^{3+}$ ions order in a triangular configuration along the basal plane [see Fig. \ref{setup}(a)], with a unique sense of rotation of moments in the triangular units. This ferrochiral arrangement is helically modulated from plane to plane with a pitch $(0,0,\tau)$, where $\tau \approx 1/7$. The balance among the inter-plane nearest and next-nearest neighbor exchanges gives a unique helicity for the spin modulation, resulting in a single helical magnetic domain \cite{Marty2008}.\par

Resonant X-ray Diffraction (RXD) experiments on BNFS found magnetic Bragg reflections along $(0, 0, n\tau) [n = 1,2,3]$\cite{Scagnoli2013}. These were attributed to out-of-plane butterfly-like sinusoidal modulation of the spins [see Fig. \ref{setup}(b)] and accompanying electron density distortions \cite{Scagnoli2013}. At the same time, due to the weak nature of higher harmonic reflections, it was unclear whether breaking of the triangular lattice symmetry or higher-order resonant terms contribute to the scattered intensity. \par

Reflections of type $(0, 0, L \pm n\tau) [n = 1,2,3]$ were also observed by polarized neutron scattering \cite{Chaix2016}. Since polarized neutrons probe magnetic moments perpendicular to the scattering wavevector, these reflections arise from a loss of the threefold rotational symmetry of spins within the basal plane \cite{Chaix2016,Toulouse2015}. It was also discovered that the single-ion magnetocrystalline anisotropy leads to subtle deviations in the rotation angle of the helix \cite{Chaix2016}. Such a \emph{bunching} of the helix explained not only the magnetic excitation spectrum but also the occurrence of a ferroelectric polarization along the in-plane directions in the absence of any external field \cite{Lee2014,Chaix2016}. \par

In this paper, we present a comprehensive RXD study on BTFS. Strong scattering signals enables us analyze the polarization of outgoing X-rays and explore in detail the energy dependence of the satellite reflections, which was not possible in case of BNFS. We further address the remaining open questions therein using our new results and first-principles calculations based on the FDMNES code \cite{Joly2001,Bunau2009}. As First-principles calculations are not yet a standard tool for RXD at $L_{2,3}$ edges of transition metals due to the complications involved, our study demonstrates how this technique can be applied to derive useful results. \par

In section II, we outline the experiments and describe the results using a simplified structure factor for the scattering process. Section III describes first-principles calculations and the major outcomes. In sections IV and V, we consolidate our understanding of the structure and symmetry of the material and list the conclusions. We provide additional information regarding the calculations in the Appendix.

\section{Resonant X-Ray Diffraction}
\subsection{Sample and Experimental Details}

The crystals were prepared and characterized as described in Ref. [\onlinecite{Marty2010}]. The crystal was cut along the trigonal [001] axis. The surface was polished and sample was subsequently annealed at $950 \degree$C in an Oxygen atmosphere for one week. The soft X-ray scattering experiments were performed using the RESOXS end-station\cite{Staub2008} at the X11MA beamline\cite{Flechsig2010} of the Swiss Light Source, Villigen, Switzerland. Linear horizontal $(\pi)$ and vertical $(\sigma)$ polarized light were focused at the sample with a spot-size of 130 $\times$ 100 $\mu m$. The monochromatized X-rays had an energy resolution of about 0.2 eV and  at the Fe $L_3$ edge.  The single crystalline sample was glued to a rotatable sample holder attached to the cold-finger of a He flow cryostat, and the sample was manually rotated in vacuum with an accuracy of $\pm 3 \degree$ for the azimuthal angle ($\Psi$) dependence. Scans along $(0,0,L)$ were performed as a function of energy and temperature, and the integrated intensity of the reflections were obtained by fitting a pseudo-Voigt function. Unless explicitly mentioned, the scans were performed with incoming $\pi$-polarized light. To analyze the polarization of scattered light, a multilayer analyzer\cite{Staub2008} was used. Scattered X-rays were measured using a standard AXUV-100 photodiode with a 400 nm thick Al filter to suppress secondary electrons.
\subsection{Results}
\begin{figure}[h] 
\centering
\includegraphics[width=0.5\textwidth]{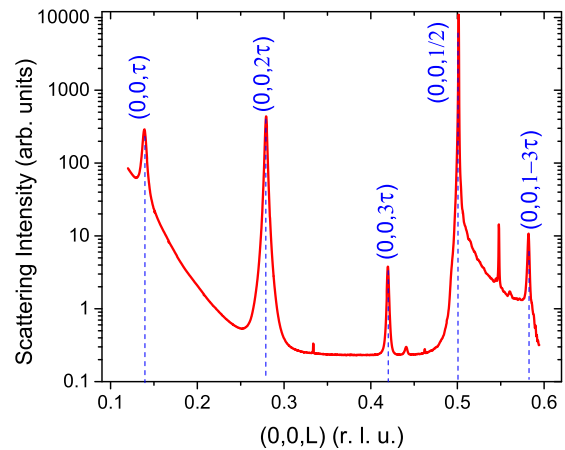}
\caption{Scan along $(0,0,L)$ direction in reciprocal space of BTFS using incoming $\pi$-polarized X-rays of energy 710 eV obtained at 10K. The peak at $(0,0,1/2)$ is the structural $(0,0,1)$ Bragg reflection observed due to $\lambda/2$ leakage from the monochromator. The peaks which are not labeled are neither resonant nor have any temperature dependence.}
\label{longscan}
\end{figure}

Resonant soft X-ray diffraction has been widely employed in recent years for the study of incommensurate magnetic systems\cite{Windsor2015,Mulders2010,Schierle2010,Shimamoto2016}. We performed detailed RXD experiments on Ba$_3$TaFe$_3$Si$_2$O$_{14}$ at the $L_{2,3}$ absorption edges of Fe and the $K$ edge of oxygen. Magnetic satellites up to the fourth harmonic fall within the Ewald sphere at the Fe $L_3$ edge, and correspondingly the satellites $(0, 0, \tau), (0, 0, 2\tau), (0, 0, 3\tau)$ and $(0, 0, 1-3\tau)$ were observed [see Fig. \ref{longscan}]. The value of the propagation vector $\tau$ was found to be $0.140 \pm 0.001$, which deviates sufficiently from (1/7) to enable us to correctly identify the peak at $L = 0.58$ as $(0, 0, 1-3\tau)$ and not $(0, 0, 4\tau)$.\par

\begin{figure}[h]
\centering
\includegraphics[width=0.35\textwidth]{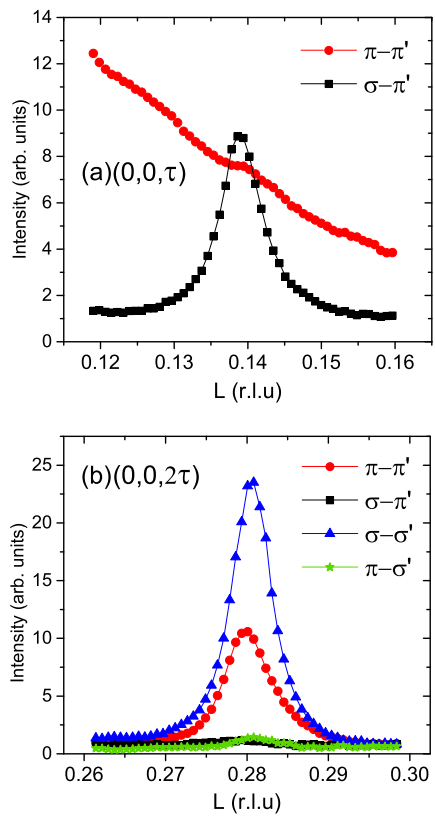}
\caption{(a) $(0, 0, \tau)$ and (b) $(0, 0, 2\tau)$ for different incoming and outgoing polarization channels for 709.8 eV at 10K. The slope in the $\pi - \pi '$ channel in (a) originates from diffuse charge scattering at small angles.}
\label{poldep}
\end{figure}

\begin{figure}[h]
\includegraphics[width=0.5\textwidth]{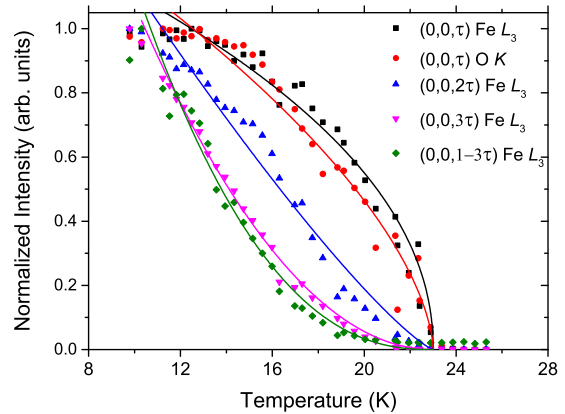}
\caption{Temperature dependence of the observed satellite reflections, with the intensities normalized to unity at 10K. The solid lines represent critical exponent fits.}
\label{tdep}
\end{figure}

\begin{figure}[h]
\includegraphics[width=0.5\textwidth]{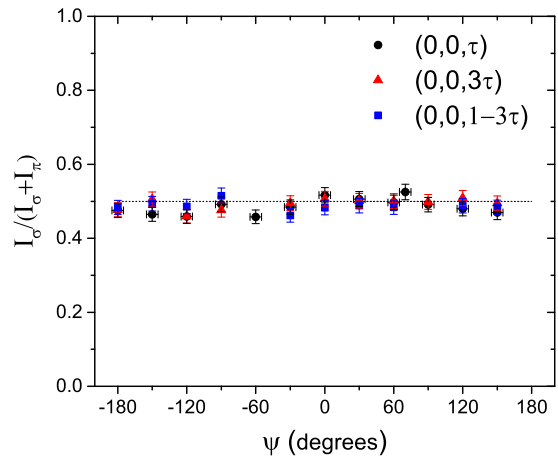}
\caption{Azimuthal dependence of the intensity ratio $I_\sigma/(I_\sigma+I_\pi)$ of the odd harmonic satellites observed at the Fe $L_3$ edge (710 eV).}
\label{azdep}
\end{figure}

The odd harmonic reflections have equal intensities for both $\sigma$ and $\pi$ incoming polarizations when the outgoing polarization is not analyzed. The $(0, 0, 2\tau)$ is stronger for $\sigma$-polarized incoming X-rays. The scattered intensities are sufficiently strong for the $(0, 0, \tau)$ and $(0, 0, 2\tau)$ to perform a polarization analysis of the scattered light [see Fig. \ref{poldep}]. The fundamental harmonic $(0, 0, \tau)$ shows a magnetic signal only in the rotated polarization channels, with the unrotated channels consisting entirely of a sloping background from diffuse charge scattering. The second harmonic $(0, 0, 2\tau)$ scatters predominantly in the unrotated channels, but one cannot rule out a weak signal in the rotated channels. These reflections exist only in the antiferromagnetic phase and show a critical exponent behavior with temperature [see Fig. \ref{tdep}] similar to that observed in case of BNFS \cite{Scagnoli2013} and certain $4f$ metals \cite{Helgesen1994,Helgesen1995}. The temperature dependence of these satellites can be fitted to: $I = I_0 (T_N - T)^{2\beta}$. The values of $\beta$ thus obtained were $0.24\pm0.06, 0.6\pm0.1, 1.0\pm0.2$ and $ 1.2\pm0.2$ for the $(0,0,\tau)$, $(0,0,2\tau)$, $(0,0,3\tau)$ and $(0,0,1-3\tau)$ taken at the Fe $L_3$ edge and $0.31\pm0.06$ for the $(0,0,\tau)$ taken at the O $K$ edge, respectively. The value of $\tau$ remains constant below $T_N$. None of the reflections show any modulation with azimuthal angle ($\Psi$) rotation within our experimental accuracy [see Fig. \ref{azdep}]. \par

\subsection{Origin of Satellites} \label{ssecorg}
To understand the nature of the satellite reflections at the Fe edge, we write a simplified structure factor - 

\begin{equation}
\label{eq3}
S(\mathbf{Q},E) = \sum_n{f_n(E) e^{i\mathbf{Q.r_n}}}
\end{equation}

where $f_n(E)$ is the energy dependent form factor of the $n^{th}$ atom located at $\mathbf{r_n}$. We only consider the scattering contribution from the Fe atoms because the observed intensity is zero away from the Fe $L_{2,3}$ edges. The expression for the resonant form factor for a dipole ($E1$) process was derived by Hannon \emph{et al.} \cite{Hannon1988}, and is given by -  

\begin{equation}
\label{eq0}
f_n = (\mathbf{\hat{\epsilon}'.\hat{\epsilon}})F^{(0)} - i(\mathbf{\hat{\epsilon}'}\times\mathbf{\hat{\epsilon}}).\mathbf{\hat{m}} F^{(1)} + (\mathbf{\hat{\epsilon}'.\hat{m}})(\mathbf{\hat{\epsilon}.\hat{m}}) F^{(2)}
\end{equation}

where $\mathbf{\hat{\epsilon}}$ and $\mathbf{\hat{\epsilon}'}$ refer to the incoming and outgoing photon polarizations, and $\mathbf{\hat{m}}$ is the unit-vector along the direction of the magnetic moment. $F^{(0)}$, $F^{(1)}$ and $F^{(2)}$ are scattering cross-sections which are complex tensors. These are derived from atomic wave-functions, and thus depend strongly on energy near atomic absorption edges.\par

We rewrite the the above equation using the formalism given by Hill and McMorrow\cite{Hill1996}. In our frame of reference, the Cartesian components of the magnetic dipole moment $m_1$ and $m_3$ lie on the horizontal scattering plane when $\Psi$ = 0, with $m_3$ parallel to the Bragg wavevector for the $(0,0,n\tau)$ reflections [see Fig. \ref{setup}]. In this setting, the resonant form factor of an Fe atom for a Bragg angle $\theta$, in $\sigma$- and $\pi$-polarized incoming X-rays can be written as - 
\begin{equation}
\begin{split}
\label{eq1}
\mid f_\sigma| =& |F^{(0)}+F^{(2)}m^2_2| + |-iF^{(1)}(m_3 sin\theta-m_1 cos\theta)\\
                 &-F^{(2)}m_2 (m_1 sin\theta+m_3 cos\theta)|
\end{split}
\end{equation}

\begin{equation}
 \begin{split}
 \label{eq2}
 \mid f_\pi| =  & |F^{(0)} cos2\theta+iF^{(1)}m_2 sin2\theta \\
                &-F^{(2)}cos^2 \theta(m^2_1 tan^2\theta+m_3^2)|\\
                &+|-iF^{(1)}(m_3 sin\theta+m_1 cos\theta)\\
					  		 &+F^{(2)}m_2 (m_1 sin\theta+m_3 cos\theta)|
 \end{split}
\end{equation}

We now use these expressions to understand the origin of the fundamental harmonic $(0, 0, \tau)$. In the high-temperature unit cell of BTFS, the three Fe atoms are crystallographically equivalent. In such a single-atom picture, the scattered intensity $I \approx |f|^2$. For the fundamental Bragg reflection $(0, 0, \tau)$, we observe that $I_\sigma \approx I_\pi$ for all values of the azimuthal angle $\Psi$. By inspection of Eqs. (\ref{eq1}) and (\ref{eq2}), the experimentally observed $I_\sigma \approx I_\pi$ can occur only if - 

\begin{equation}
\label{eq4}
f_\sigma \approx f_\pi \approx -iF^{(1)} m_3 sin\theta
\end{equation}

This means that the charge and orbital terms ($F^{(0)}$ and $F^{(2)}$) are negligible compared to $F^{(1)}$, which is expected for a magnetic satellite reflection. An in-plane spin moment contributing to the satellites has been observed by neutron scattering\cite{Chaix2016}. Within our experimental precision, we cannot rule out a contribution from the spin moments $m_1$ and $m_2$. From the marginal deviation in the average value of $I_{\sigma}/(I_{\pi}+I_{\sigma})$ ($0.486 \pm 0.006$ instead of $0.5$ for $(0, 0, \tau)$, for example), we can estimate an upper limit of 0.2 in the ratio of $m_1/m_3$ (or equivalently $m_2/m_3$). But the absence of a clear azimuthal dependence implies that it is unlikely that $m_1$ and $m_2$ contribute significantly to any of the reflections. Hence, only the contribution of $m_3$ which is along the $\mathbf{\hat{c}}$-axis (given by Eq. (\ref{eq4})), is considered for further discussions on the $(0,0,\tau)$ reflection.\par

The second harmonic $(0, 0, 2\tau)$ arises from electron density distortions created by the magnetic helix, contained in the terms $F^{(0)}$ and $F^{(2)}$ of Eqs. (\ref{eq1}) and (\ref{eq2}). The ratio of $I_\sigma/I_\pi$ depends strongly on energy, similar to that observed in BNFS\cite{Scagnoli2013}. So this reflection can have more than one nonzero term in the form factor, due to which a detailed analysis becomes non-trivial.\par 

The third harmonic satellites $(0, 0, 3\tau)$ and $(0, 0, 1-3\tau)$ can have many different origins. In some magnetic oxides such as CuB$_2$O$_4$, strong spin-lattice coupling and competing interactions cause soliton-like distortions in the spin structure, leading to additional Fourier components in reciprocal space\cite{Zheludev1997,Roessli2001,Togawa2012,Narita2016}. In rare-earth metals that have a $\mathbf{\hat{c}}$-axis spiral spin structure\cite{Helgesen1995}, the third harmonic originates from higher order resonant processes. In contrast to the rare-earth metals\cite{Takai2015}, presence of higher-rank multipoles is unlikely for $3d$ metals, and hence were neglected in Eqs. (\ref{eq1}) and (\ref{eq2}). For example, a quadrupole transition at the $L_3$ edge involves the transition of an electron from the $2p_{3/2}$ state to the unoccupied $4f$ levels of Fe. These levels are part of the continuum and are devoid of any information related to magnetism, to a good approximation. Hence, these contributions are expected to be very weak. In our case, the polarization and azimuthal angle dependence of the third harmonic satellites are the same as that of the fundamental harmonic $(0, 0, \tau)$. This suggests that the $(0, 0, 3\tau)$ and $(0, 0, 1-3\tau)$ appear due to deviations from the sinusoidal spin modulation along the $\mathbf{\hat{c}}$-axis, and is described by the same component as in Eq. (\ref{eq4}). \par

\subsection{Energy Dependence at Fe $L_{3}$ Edge}
\label{ssecedep}

\begin{figure} [h]
\includegraphics[width=0.5\textwidth]{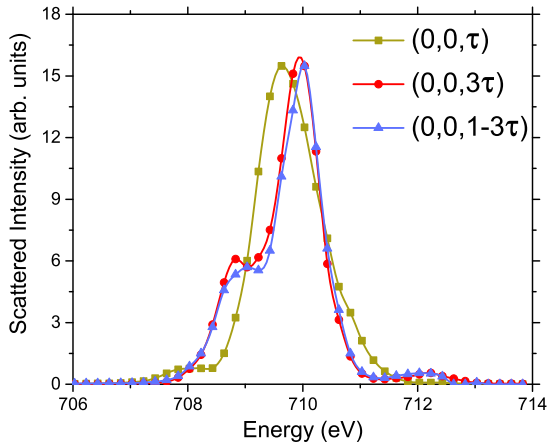}
\caption{Energy dependence of integrated intensity of the odd harmonic satellites at the Fe $L_3$ edge. The points on the figure correspond to integrated intensity of a scan along $(0, 0, L)$ at the corresponding energy (normalized to the peak intensity of $(0, 0, \tau)$ for comparison). The solid lines are a guide to the eye.}
\label{edep}
\end{figure}

To gain more insight into the nature of these satellites, we compare their energy dependences. The energy dependences of the odd harmonic reflections in the vicinity of the Fe $L_3$ edge are shown in Fig. \ref{edep}. Each point on the spectrum corresponds to the integrated intensity of the Bragg peak obtained from a scan along $(0,0,L)$ for the given energy. There is a clear difference in the shapes of the first and third harmonic satellites. The intensities have been corrected for self-absorption. By performing a reciprocal space scan along the scattering wavevector, refraction effects have been accounted for. The similarity between $(0, 0, 3\tau)$ and $(0, 0, 1-3\tau)$ further supports the argument that the observed differences are not influenced by geometrical parameters like difference in Bragg angles, but are rather intrinsic to the order of the satellite reflection. In the following paragraphs, we explore the cause of this difference in the energy dependence using the expressions derived in the previous section [Eqs. (\ref{eq3})-(\ref{eq4})]. \par

In the expression for the structure factor [Eq. (\ref{eq3})], only the phase term $e^{i\mathbf{Q.r_n}}$ depends on $\mathbf{Q}$. The form factor $f_n(E)$ consists of a single resonant term [see Eq. (\ref{eq4})]. Hence, the energy dependence of $S(\mathbf{Q})$ should be independent of $\mathbf{Q}$ for an incommensurate reflection. In this case, $(0, 0, \tau)$ and $(0, 0, 3\tau)$ should have an identical energy dependence. However, this is not true if there exists a fixed relationship between $f_n(E)$ and $e^{i\mathbf{Q.r_n}}$. In BTFS, the cases where this can occur are (i) emergence of non-equivalent Fe atoms in the unit-cell because of breaking of the threefold lattice symmetry, (ii) modification of local electronic structure by the spins leading to structural incommensurability, and (iii) intrinsic difference in lineshape arising due to quantum mechanical interference of scattering amplitudes from different core-hole states. Cases (i) and (ii) are interrelated. However, for better understanding of their influence on the spectral shapes, we discuss them independently in the following paragraphs.\par



To understand case (i), we look at the high-temperature unit cell of BTFS, which contains three symmetry equivalent Fe atoms related by a simple rotation. However, in presence of Dzyaloshinskii-Moriya interactions, the atoms are displaced from their equilibrium positions\cite{Walker2011} lowering the crystal symmetry. The magnitude of these displacements depend on the net spin-moment on the atoms, and thus, the phase factor $e^{i\mathbf{Q.r_n}}$ of the $n^{th}$ atom becomes dependent on the corresponding magnetic form factor $f_n$. The inverse effect can also occur where $f_n$ is modified due to a displacement $\delta r_n$. In presence of such displacements, Eq. (\ref{eq3}) can be rewritten as -
\begin{equation}
\label{eq6}
S_{\sigma,\pi}(\mathbf{Q},E) = \sum_n{(f_n + \delta f_n) e^{i\mathbf{Q.(r_n+\delta r_n)}}} 
\end{equation}
where $\delta f_n$ is the change in $f_n$ due to the displacement $\delta r_n$. Now the scattered waves add up differently as a function of energy for different $\mathbf{Q}$-vectors. In order to change the phase of the scattered wave for $(0, 0, \tau)$ or $(0, 0, 3\tau)$, the displacement should have a component along the $\mathbf{\hat{c}}$ axis. Any displacement of this type which is consistent with the observed electric polarization\cite{Lee2014} breaks the threefold symmetry of the lattice. This also introduces crystallographic non-equivalence of the Fe atoms in the unit-cell. \par

With regard to case (ii), the expressions [Eqs. (\ref{eq1})-(\ref{eq4})] for resonant amplitudes of the magnetic form factor $f_n$ hold only under spherical symmetry. Magnetocrystalline anisotropy increasing tendency of spins to preferentially align along easy-planes (perpendicular to the twofold local axes) has been observed by neutron scattering\cite{Chaix2016}. Conversely, if the crystal structure follows the spin orientation, this might result in asphericity of part of the electron cloud\cite{Haverkort2010}. Thus, it is possible to have a modulation of the local electronic environment whose periodicity depends on the magnetic structure. Difference in spectral shapes for dissimilar Bragg reflections, is now a straightforward outcome. In this case, the crystal structure can no longer be considered commensurate. The appearance of the $(0, 0, 2\tau)$ reflection due to deformation of the Fe $3d$ orbitals is a direct evidence of an incommensurate structural modulation. The helical electric polarization observed by terahertz spectroscopy\cite{Chaix2013} also supports this picture. Yet another observation in this regard is that the value of the magnetic modulation wavevector $\tau$ remains constant for all temperatures, indicating a strong coupling to the lattice.

As for case (iii), the splitting of core-levels leading to the hyperfine structure of spectral lines is a well-known phenomenon in atomic physics and core-level spectroscopies. In the context of RXD, effects of such a splitting have been discussed for the case of $M_{4,5}$ edges of $4f$ atoms \cite{Mulders2006,Princep2011}. In BTFS, due to the non-collinear spin structure, the axis of relativistic spin quantization varies from one atom to the next. Even though visible effects due to this phenomenon are rather uncommon for the $3d$ elements, we include a brief description in the Appendix.\par

It is important to note that all the above distortions (cases (i)-(iii)) affect only the immediate environment of the atom and thus can be much better detected in diffraction experiments with X-rays tuned to strong atomic absorption resonances. In the following section, we make use of first-principles calculations to demonstrate cases (i) and (ii) discussed above.\par 

\section{First-principles Calculations}
\label{secfdmnes}

The Fe $L_3$ edge has a multiplet structure and therefore, density functional theory (DFT) based methods are not very precise in calculating the spectral function\cite{Bunau2012}. However, this method is effective to qualitatively examine how changes in the crystal structure affect the spectral shapes of the satellites in relation to one another. We use the FDMNES\cite{Joly2001,Bunau2009} package to calculate the energy dependences of the different reflections. The program uses a given crystal structure and a starting magnetic structure to compute the spin-polarized electronic density of states of an absorbing atom, and subsequently calculates diffraction spectra for well-defined X-ray polarizations and Bragg wavevectors. Furthermore, the program performs a complete multipolar analysis\cite{Matteo2012} for any given X-ray absorption process. To approximate the incommensurate magnetic ground-state, we create a supercell consisting of seven conventional unit-cells stacked on top of each other along the trigonal $\mathbf{\hat{c}}$-axis, with a periodicity close to the observed magnetic modulation wavevector $(0, 0, \tau)$. We place the spin moments on Fe atoms in a $120 \degree$ triangular lattice, which rotates in a perfect helix with a pitch that equals the $c$ lattice parameter of the supercell. 

\subsection{Calculations for the Fe $L_3$ edge}
\label{ssecfdfe}
The validity of the supercell approximation described above is verified by the fact that the $\mathbf{\hat{c}}$-axis dipole moments emerge automatically in the calculation as a result of reconstruction of spin-polarized local density of states on individual atoms. The dominant calculated intensity for the fundamental reflection $(0, 0, \tau)$ appears in the rotated polarization channels and the ratio $I_\sigma/(I_\sigma+I_\pi) = 0.5$ is in excellent agreement with our experiments. A multipole expansion of the scattered signal using spherical tensors agrees with the fact that the intensity is nonzero only for the $\mathbf{\hat{c}}$-axis component of the magnetic dipole moment [Eqs. (\ref{eq3})-(\ref{eq4})].\par

\begin{figure}[h]
\centering
\includegraphics[width=0.45\textwidth]{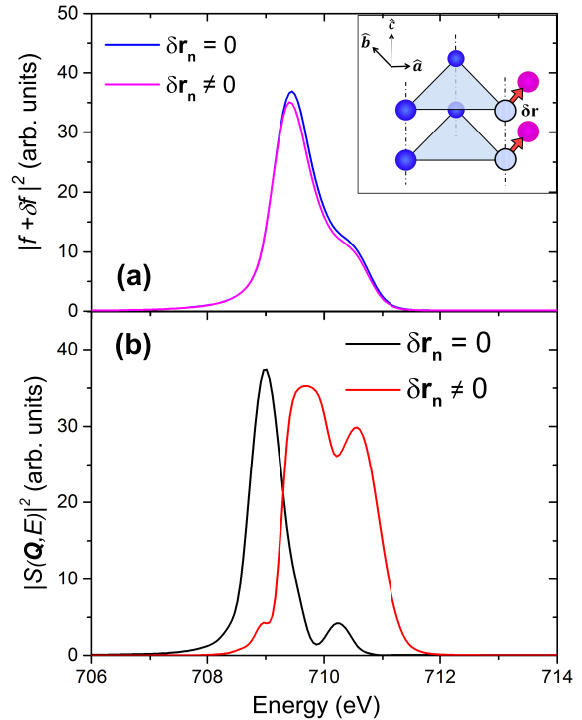}
\caption{(a) Spectral contribution of the magnetic form factor for the undisplaced and displaced atoms, and (b) spectral shape of the $(0,0,3\tau)$ Bragg reflection, at Fe $L_3$ edge. The inset of (a) shows schematically how the atoms are displaced simultaneously along both in-plane and out-of-plane directions. The intensities in (b) have been rescaled for comparing the shapes.}
\label{fdfe}
\end{figure}

This technique can now be used to study how the energy dependences are modified by small displacements of the Fe atoms in the crystal structure. A net electric polarization has been observed along both $\mathbf{\hat{a}}$ and $\mathbf{\hat{c}}$ directions\cite{Lee2014}. In order to be consistent with this observation, the displacements should break the local twofold axis at the atomic sites as well as the threefold symmetry of the unit-cell. The simplest way to achieve this would be to displace one Fe atom within each triangular unit of the supercell both along $\mathbf{\hat{a}}$- and $\mathbf{\hat{c}}$-axes [see inset of Fig-\ref{fdfe} (a)]. We displace the atoms by an arbitrary number of $0.03\AA$ in our calculations. In principle, one could choose to displace any one or more of the Fe atoms along any arbitrary direction such that they are consistent with the observed polarization. At first, we look at the change in the energy dependence of the magnetic form factor due to this displacement. Figure-\ref{fdfe} (a) shows the spectral intensity profile resulting from the magnetic scattering factor at the Fe $L_3$ edge for undisplaced and displaced atoms. The spectral contribution of the displaced atom is marginally different from the undisplaced atom due to its modified electronic environment [described by Eq. (\ref{eq6})]\par

Next, we examine the resulting change in the energy dependence of a magnetic Bragg reflection. We choose the $(0, 0, 3\tau)$ at the Fe $L_3$ edge as an example. To eliminate computational errors that may occur due to the supercell approximation, we perform this calculation by the selective use of scattering terms described in Eq. (\ref{eq6}) to calculate the structure factor $S_{\sigma,\pi}(\mathbf{Q},E)$ for the supercell. Figure-\ref{fdfe}(b) shows the change in energy dependence of the $(0, 0, 3\tau)$ reflection resulting from the atomic displacements described above. The significantly different energy dependence for the $(0,0,3\tau)$ reflection is the result of strong phase shifts due to atomic displacements, which are enhanced considerably at absorption resonances. \par

\begin{figure}[h]
\centering
\includegraphics[width=0.45\textwidth]{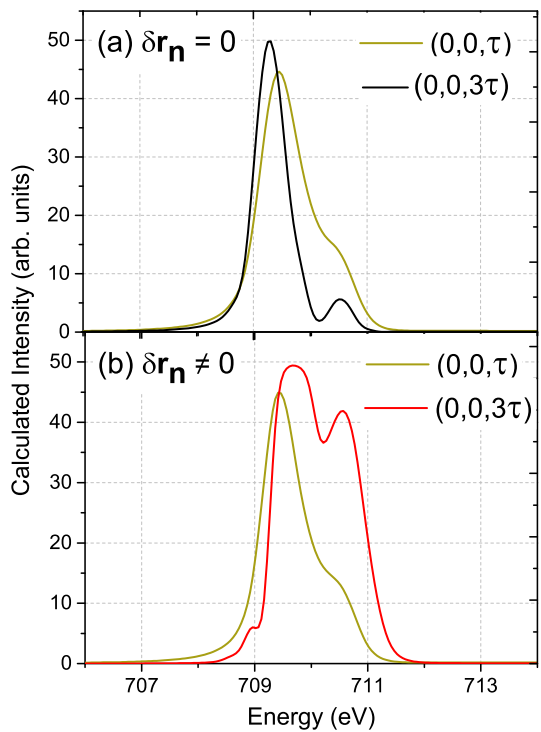}
\caption{Calculations of the spectral intensity at the Fe $L_3$ edge for the $(0,0,\tau)$ and $(0,0,3\tau)$ reflections for (a) three equivalent Fe atoms, and (b) one of the three atoms displaced within the unit cell of BTFS. The intensities have been rescaled to enable comparison of the shapes.}
\label{fdfe2}
\end{figure}

We are now in a position to compare the energy dependence of $(0,0,\tau)$ and $(0,0,3\tau)$ reflections when an atom is displaced. Figure \ref{fdfe2}(a) shows the spectra of the two Bragg reflections for the case of equivalent undisplaced Fe atoms, where the local structure and spins are solely rotated by 120$\degree$ between any two given atoms. In this case, both spectra show two features at very similar energies [see Fig. \ref{fdfe2}(a)]. In the case when one of the atoms is displaced and no longer symmetry equivalent, the resulting spectra have new and distinct features [see Fig. \ref{fdfe2}(b)]. The enhancement or suppression of the different spectral features are extremely sensitive to the nature and magnitude of the atomic displacements used in the calculation. Hence, one cannot compare their intensities in absolute terms. But a relative difference in shape of $(0,0,\tau)$ and $(0,0,3\tau)$ is always observed, as long as the atomic displacements are consistent with the electric polarization. This demonstration is the essence of the relationship between the magnetic form factor and the phase term discussed in the previous section (cases (i) and (ii)), and indicates the emergence of inequivalent Fe sites in BTFS below $T_N$. Since the phase shift term is a product of the $\mathbf{Q}$ and $\mathbf{\delta r_n}$, the changes are more pronounced for the $(0,0,3\tau)$ due to the larger value of $|\mathbf{Q}|$ [see Eq. (\ref{eq6})]. This also shows that the subtle deviation from the sinusoidal magnetization profile is more sensitive to symmetry changes in the system.\par


On repeating such calculations for several types of displacements, we find that only those which break the threefold crystal symmetry produce significant differences in spectral shapes. One should also point out that for the case of equivalent atoms, the energy dependence of $(0,0,\tau)$ and $(0,0,3\tau)$ is not the same even though the spectral features appear at the same energies. This is due to the interference of core-hole states, discussed in more detail in the Appendix.\par

\subsection{Calculations for the O $K$ edge}
\begin{figure}[h]
\includegraphics[width=0.45\textwidth]{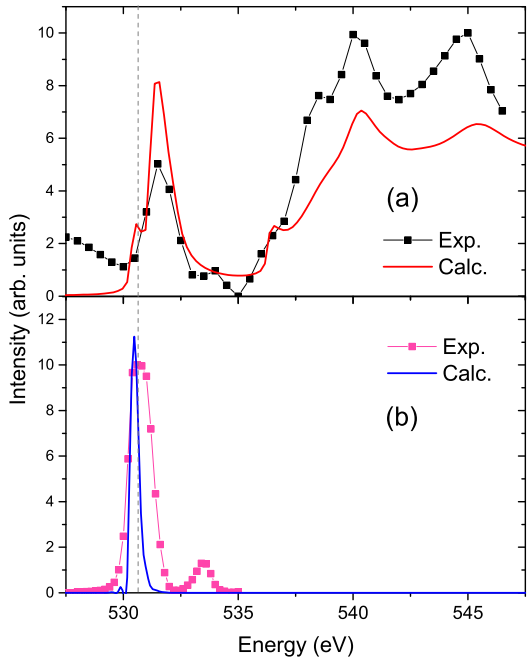}
\caption{Comparison of experimentally observed and calculated (a) fluorescence spectrum at the Oxygen $K$ edge, and (b) $(0,0,\tau)$ magnetic satellite. The solid lines connecting the experimental data points are guides to the eye.}
\label{fdox}
\end{figure}

The calculated energy dependences shown in Fig. \ref{fdfe2} do not contain all features present in the experimentally observed spectra [Fig. \ref{edep}]. This is a fundamental drawback of DFT-based methods applied to transition metal $L_{2,3}$ edges which contain atomic multiplets due to the localization of $3d$ states in presence of a core-hole\cite{Bunau2012}. These limitations do not apply to the $K$ edge due to the $s$-character of the core state. To demonstrate this, we use the same supercell and reference magnetic structure to calculate the absorption and diffraction spectra for the oxygen $K$ edge of BTFS. Figure \ref{fdox} shows the experimental and calculated spectra at the oxygen $K$ edge. The most prominent diffraction signals coincide with the rising half of the pre-edge absorption feature which belongs to the oxygen $2p$ levels which are strongly hybridized with the Fe $3d$ orbitals. Such a strong hybridization results in partial transfer of magnetic moment to the oxygen atoms, a phenomenon that has been observed also in other systems\cite{Beale2010,Mannix2001,Weht1998,Huang2016}. The temperature dependence of this reflection is similar to the one observed at the Fe $L_3$ edge, further indicating a purely magnetic origin of this peak [see Fig. \ref{tdep}]. The calculated intensity for the $(0,0,2\tau)$ satellite is zero. Even though the calculation produces a non-zero $(0,0,3\tau)$ reflection, its calculated intensity (about 100 times weaker compared to the calculated $I^{(0,0,\tau)}$) suggests that it might be too weak to be experimentally detected.

Since the core-state in case of a $K$ edge is not spin-orbit split, the magnetic dipole probed by an $E1$-$E1$ transition at this edge represents the occurrence of an orbital magnetic moment on the oxygen atoms\cite{Lovesey2005}. Hence, the diffraction signal at $(0,0,\tau)$ appears only in presence of spin-orbit coupling of the valence states as confirmed by the calculations.\par


\section{Discussion}
This resonant soft X-ray diffraction study on BTFS sheds light on the low temperature crystal symmetry of Fe containing multiferroic langasites. Both BTFS and BNFS have a nearly identical spin structure and a similar value of $T_N$ ($\approx27K$) below which we observe the satellite reflections $(0,0,\tau)$, $(0,0,2\tau)$, $(0,0,3\tau)$ and $(0,0,1-3\tau)$ at the $L_{2,3}$ edges of Fe and the $(0,0,\tau)$ at the $K$ edge of oxygen. The odd harmonics are dominated by the butterfly magnetic component, i.e., the canting of the magnetic dipoles along the trigonal $\mathbf{\hat{c}}$-axis, while the $(0,0,2\tau)$ arises due to the electron density distortions accompanying magnetic order. The dependence of these satellites on temperature, X-ray polarization and azimuthal angle are the same within experimental accuracy, for both the compounds. We do not see any direct effect of the breaking of the triangular spin configuration in the satellites, since they are dominated by the $\mathbf{\hat{c}}$-axis butterfly magnetic modulations [Eqs. (\ref{eq0})-(\ref{eq4})]. The microscopic interactions, however, differ between BTFS and BNFS\cite{Chaix2016} causing observable changes in the experiment. For example, the value of $\tau$ differs significantly for the two compounds. We observe stronger scattering intensities for the higher order satellites in BTFS compared to BNFS. Since the $(0,0,3\tau)$ and $(0,0,1-3\tau)$ are a measure of deviation from the sinusoidal modulation, larger scattering intensities can be a consequence of stronger single-ion anisotropy in BTFS\cite{Chaix2016}.\par

Lee \emph{et al.}\cite{Lee2014} observed a spontaneous electric polarization in BNFS with the onset of magnetic order, along both the trigonal $\mathbf{\hat{a}}$ and $\mathbf{\hat{c}}$-axes. For an atomic displacement that creates this electric polarization, the threefold symmetry of the crystal can be lost, leading to the emergence of non-equivalent Fe sites within the structure. So far, no conclusive evidence of any atomic displacements have been reported, even though several experiments indicate a lowering of symmetry\cite{Lyubutin2011,Chaix2013,Toulouse2015}.\par

We observe differences in the energy dependences between the first and third order satellite reflections, which are directly related to magnetism. First-principles calculations show that these differences can be qualitatively understood in terms of phase shifts due to displacement of the Fe atoms. Emergence of non-equivalent Fe sites within each triangular unit of the structure is necessary to obtain different spectral shapes for the $(0,0,\tau)$ and $(0,0,3\tau)$ reflections. Thus, we observe a breaking of threefold crystal symmetry indirectly from magnetic scattering. Our results complement the recent observations made by neutron scattering\cite{Chaix2016}. Further research is needed to understand the exact symmetry in the antiferromagnetic phase and its implications on the interactions driving multiferroicity.\par 

It is well-known that RXD is sensitive to a variety of electronic ordering phenomena, and the energy dependence contains a wealth of information on the electronic and magnetic structure. Considering the fact that the information contained in $L_{2,3}$ edge diffraction spectra are seldom examined for Bragg scattering (especially for partially filled $d$- and $f$-electron systems), our demonstration of extracting electronic and magnetic information contained in the spectra is significant. Effects emerging from strong electron correlations, artefacts due to self-absorption and core-hole effects make quantitative evaluation of spectral lineshape at the $L (M)$ edges of partially filled $d$-orbital ($f$-orbital) systems extremely challenging. Recently, there has been some progress made in tackling dynamical effects in the soft X-ray regime\cite{Macke2016}. However, few computational codes are available that incorporate multi-electronic wave functions in multi-atomic systems. Therefore, a comprehensive analysis of the spectral features is presently not straightforward and therefore ignored for most cases. We limit ourselves to systematic simulation of the observed experimental energy dependence of the different reflections, and use this to selectively examine effects of symmetry lowering in the system. Even though such an approach is neither exhaustive nor universal, it provides valuable insights into the diffraction process at an atomic level for the case of BTFS. In any case, further work is needed in improving the applicability of computational codes to enable better understanding of RXD spectra, in particular in the regime where band and local effects are both important.\par

\section{Conclusions}
We have studied the chiral multiferroic langasite Ba$_3$TaFe$_3$Si$_2$O$_{14}$ using resonant soft X-ray diffraction at the Fe $L_{2,3}$ edges and the oxygen $K$ edge. The difference in the spectral shapes of magnetic satellites $(0,0,\tau)$ and $(0,0,3\tau)$ can be attributed to atomic displacements due to magnetoelastic coupling in the material. This supports the loss of threefold crystal symmetry in the antiferromagnetic state of the system, as shown by the first-principles calculations. Finally, we demonstrate how one can utilize first-principles calculations to understand structure and symmetry information obtained in X-ray magnetic scattering on complex systems.

\section*{Acknowledgements}
We thank S. W. Lovesey, A. Scaramucci and J.M. Perez-Mato for stimulating discussions. The experiments were carried out at X11MA beamline of the Swiss Light Source, Paul Scherrer Institut, Villigen, Switzerland. The authors thank the X11MA beamline staff for experimental support. The financial support of the Swiss National Science Foundation (SNSF) is gratefully acknowledged. E.M.B. acknowledges financial support from NCCR Molecular Ultrafast Science and Technology (NCCR MUST), a research instrument of the SNSF, and funding from the European Community’s Seventh Framework Program (FP7/2007-2013) under Grant Agreement No. 290605 (COFUND: PSI-FELLOW).

\appendix
\section{FDMNES - Approximations and Limitations}
The crystal structure used in the calculations is the same as the one described in Ref. [\onlinecite{Marty2010}]. The supercell constructed for the calculations consists of seven conventional unit cells stacked along the trigonal $\mathbf{\hat{c}}$-axis. The number seven is chosen since the value of the modulation wavevector $\tau (\approx 0.14)$ is close to $1/7$. The input magnetic structure consists of spins in a perfect triangular configuration (with no $\mathbf{\hat{c}}$-axis component), rotating in a perfect helix having a pitch equal to the $\mathbf{\hat{c}}$ lattice constant of the supercell.\par

For a nonmagnetic calculation on this supercell, we obtain zero intensity for all the satellite reflections. For a magnetic calculation, we find non-zero intensities for $(0,0,\tau)$, $(0,0,2\tau)$, $(0,0,3\tau)$ and $(0,0,1-3\tau)$ at the $L_{2,3}$ edges of Fe. The scattering signal for the odd-harmonic satellites is dominated by the $\mathbf{\hat{c}}$-axis magnetic component $m_3$, and we do not find any contribution of the in-plane magnetic component $m_1$ or $m_2$. The relative intensities obtained in the calculations match the experiment, correct to an order of magnitude. \par

Even though the value of the propagation vector $\tau$ is very close to $1/7$, a supercell approximation creates computational errors. For the odd harmonic reflections, we find intensities in the unrotated polarization channels, unlike what is observed experimentally. A detailed analysis reveals that this intensity comes from charge and orbital density reconstruction in presence of magnetic order and relativistic interactions. Hence, they can be safely ignored for the purpose of this study. Since the calculated intensities of resonant scattering may be inaccurate, even weak non-resonant magnetic scattering signals can interfere with the resonant part giving rise to artefacts in the energy dependence. Hence, we have forced the non-resonant magnetic scattering to zero for the calculations described in Sec. \ref{ssecfdfe}. \par

For comparison of spectral shapes, we take the spherical harmonic expansion of the scattering tensor of each atom and build the structure factors for the relevant Bragg reflections using only the relevant part of the scattering tensor. Since this method eliminates errors due to the supercell approximation which mainly appear as charge (not magnetic) scattering, we can use it to compare the spectral shapes originating purely from magnetic scattering. Following the discussion about high-order multipoles in Sec. \ref{ssecedep}, we checked for the presence of scattering signals from the $E2-E2$, $E1-E2$ and $E1-M1$ processes and found them to be negligible. Hence, we confine ourselves to the $\mathbf{\hat{c}}$-axis magnetic component derived from the dipole $E1-E1$ scattering term for all calculations described in Sec. \ref{secfdmnes}.\par

The calculations were done using a cluster radius of 3.8 $\AA$, which includes the nearest-neighbor Fe atoms. A uniform broadening of 0.15 eV was included. Surprisingly, we found that the multiple scattering (MS) approach on a muffin-tin potential gave better results than the finite difference method (FDM). We can conjecture that the use of a grid of points on a fixed supercell in FDM, which is different from reality, leads to more errors than the muffin-tin approximation does within the MS theory.

\section{Core-resolved Spectral Intensities}
A dipole absorption event at the $L_3$ edge of Fe involves the creation of a core-hole in the $2p_{3/2}$ levels, which consist of four distinct states with $j_z$ values $-3/2$, $-1/2$, $1/2$ and $3/2$. Including the interference of waves from different core-hole states, the structure factor for the magnetic reflections can be rewritten as –


\begin{equation}
\label{eqcore}
{S}(\mathbf{Q},E) =\sum_n{ {\Big[ \sum_{j_z}{{f}_{j_z}} \Big] }_n e^{i\mathbf{Q.r_n}}} 
\end{equation}

where $f_{j_{z}}$ is the magnetic form factor corresponding to the core-hole state $j_z$. Relative differences in contribution from each state to the Bragg reflections can now account for the difference in their spectral shapes.\par

\begin{figure}[h]
\includegraphics[width=0.45\textwidth]{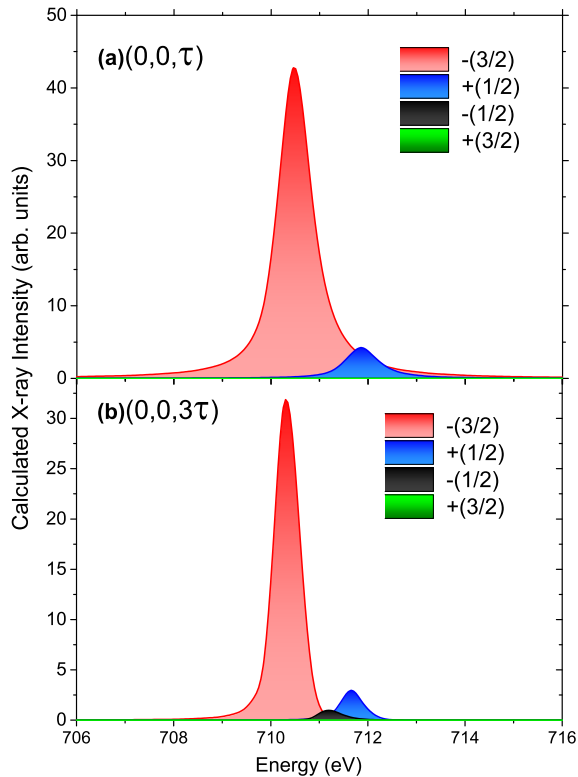}
\caption{Core-resolved spectral intensities of (a) $(0,0,\tau)$ and (b) $(0,0,3\tau)$  at the Fe $L_3$ edge. The peaks represent the spectral intensity contribution from individual $j_z$ levels of the $2p_{3/2}$ core state.}
\label{core}
\end{figure}

The calculated core-resolved spectra for the $(0,0,\tau)$ and $(0,0,3\tau)$ satellites with the transitions involving individual $j_z$ core-states are shown in Fig. \ref{core}. The $(0,0,3\tau)$ has different matrix elements connecting the $2p_{3/2}$ core state to the magnetic $3d$ states of Fe compared to the $(0,0,\tau)$, leading to differences in the final scattering spectrum between the two reflections. A detailed examination of the above interference phenomena is very interesting from a fundamental point of view, which remains to be studied in more detail in the future.

\bibliographystyle{unsrt}
\setcitestyle{numbers}
\bibliography{Jabrefs}

\end{document}